\documentclass{emulateapj}
 
\shorttitle{Analysis of K2 Short-cadence Photometry of Qatar-2b}
\shortauthors{Dai et al.}

\usepackage{amsmath}
\usepackage{epstopdf}
\usepackage{apjfonts}
\usepackage{xspace}
\usepackage{hyperref} 
\usepackage{bm}
\usepackage{graphicx}
\begin{document}


\title{The stellar obliquity, planet mass, and very low albedo of Qatar-2
  from {\it K2} Photometry}


\author{Fei\ Dai\altaffilmark{1}, Joshua N.\ Winn\altaffilmark{1,2}, Liang Yu\altaffilmark{1}, Simon Albrecht\altaffilmark{3}}


\altaffiltext{1}{Department of Physics and Kavli Institute for
  Astrophysics and Space Research, Massachusetts Institute of
  Technology, Cambridge, MA 02139, USA {\tt fd284@mit.edu}}

\altaffiltext{2}{Department of Astrophysical Sciences, Peyton Hall, 4
  Ivy Lane, Princeton, NJ 08540 USA}

\altaffiltext{3}{Stellar Astrophysics Centre, Department of Physics
  and Astronomy, Aarhus University, Ny Munkegade 120, DK-8000 Aarhus
  C, Denmark}


\begin{abstract}
  \noindent
  
The Qatar-2 transiting exoplanet system was recently observed by the
{\it Kepler} as part of {\it K2} Campaign 6 in short-cadence mode.
We identify dozens of starspot-crossing events, when the planet
eclipsed a relatively dark region of the stellar photosphere. The
observed patterns of these events demonstrate that the
planet always transits over the same range of stellar latitudes, and
therefore that the stellar obliquity is less than about 10$^\circ$.
We support this conclusion with two different modeling approaches: one
based on explicit identification and timing of the events, and the
other based on fitting the light curves with a spotted-star model.  We refine the transit parameters and measure the
stellar rotation period ($18.5 \pm 1.9$~days), corresponding to a
'gyrochronological' age of $1.4 \pm 0.3$ Gyr.  Coherent flux
variations with the same period as the transits are well modeled as the
  combined effects of ellipsoidal light variations ($15.4 \pm
  4.8$~ppm) and Doppler boosting ($14.6 \pm 5.1$~ppm). The magnitudes
  of these effects correspond to a planetary mass of $2.6 \pm
  0.9~M_{\text{Jup}}$ and $3.9 \pm 1.5~M_{\text{Jup}}$,
  respectively. Both of these independent mass estimates agree with
  the mass determined by the spectroscopic Doppler technique ($2.487
  \pm 0.086~M_{\text{Jup}}$).  No occultations are detected, giving a
2$\sigma$ upper limit of 0.06 on the planet's visual geometric
albedo.  We find no evidence for orbital decay,
although we are only able to place a weak lower bound on the relevant
tidal quality factor: $Q'_\star > 1.5\times 10^4$~(95\% confidence).

\end{abstract}


\section{Introduction}

The obliquity of a planet-hosting star (the angle between the star's
rotation axis and orbit normal) may bear information about a planet's
formation, migration and tidal evolution history \citep{Queloz2000,
  Ohta2005, Gaudi2007, Winn2010}. For example, dynamically hot
scenarios for hot Jupiter formation, such as planet-planet scattering
\citep{Chatterjee2008} and Kozai-Lidov mechanism \citep{Fabrycky2007},
should often produce large obliquities.  Dynamically cold scenarios
such as disk migration \citep{Lin1996} and {\it in situ} formation
\citep{Batygin2015} should preserve low obliquities, unless there
are mechanisms for exciting obliquities independently of hot-Jupiter
formation \citep[e.g.,][]{Bate2010, Batygin2012}.

One way to determine the stellar obliquity --- or, to be more precise,
to recognize when the obliquity is low --- is to observe a sequence of
flux anomalies that occur when a transiting planet repeatedly passes
in front of a starspot. The analysis of these ``starspot-crossing
anomalies'' takes advantage of the precise time-series photometry that
is available for the systems that have been observed by the {\it
  Kepler} and {\it CoRoT} spacecraft.  This method does not require
intensive time-series spectroscopy, unlike the more traditional method
based on the Rossiter-McLaughlin effect, which is often difficult to
apply to relatively faint or slowly-rotating stars.

\citet{Silva2003} anticipated the observable signal of a transiting
planet crossing over a starspot: the loss of light is temporarily
reduced, because the starspot has a lower intensity than the
surrounding photosphere. This produces a brief flux enhancement or
``bump'' in the transit light curve.  It soon became clear that
spot-crossing anomalies can be used to study the properties of
starspots \citep{Silva-Valio2010}, demonstrate the presence of active
latitudes \citep{Sanchis-Ojeda2011Hat} and constrain the stellar
obliquity \citep{Sanchis-Ojeda2011Wasp,Nutzman2011}.

Qatar-2b is a hot Jupiter with a mass of 2.5~$M_{\text{Jup}}$, a
radius of 1.1~$R_{\text{Jup}}$, and an orbital period of 1.34~days.
It was discovered by the Qatar Exoplanet Survey
\citep[QES,][]{Bryan2012}. The host star Qatar-2A is a relatively
bright K dwarf ($V = 13.3$, $M_\star = 0.740 \pm 0.037
~M_{\odot}$). Radial velocity follow-up revealed the presence of a
long-term trend which was attributed to a more distant
companion. \citet{Mancini2014} constrained the obliquity of Qatar-2b
using spot-crossing anomalies seen in the ground-based multi-color
transit observations. However, the stellar rotation period was unknown
at the time of their analysis. Without the ability to calculate the
rotational phase of each transit, \citet{Mancini2014} had to make the
assumption that two particular spot-crossing anomalies they observed
were caused by eclipses of the same spot. With this assumption, they
found the stellar rotation period to be  $14.8 \pm0.3$ days
  \citep[after the correction described by] []{Mancini2016}, and the
sky-projected obliquity (the angle between the sky projections of the
stellar rotation axis and the orbit normal) to be $\lambda = 4.3 \pm
4.5^{\circ}$.

Qatar-2 was within the field of view of {\it K2} Campaign 6. Being a
confirmed planet, Qatar-2 was selected for 1~min (``short-cadence'')
time sampling, instead of the usual 30~min sampling. The precise,
continuous and well-sampled {\it K2} photometric data provides an
opportunity to study Qatar-2b in greater detail. As we will show, the
{\it K2} data reveal the stellar rotation period to be $18.5 \pm 1.9$
days, at odds with the period determined by
\citet{Mancini2014}. Moreover, the {\it K2} data show evidence for
numerous spot-crossing anomalies caused by different spot groups.
This leaves little room for doubt in the interpretation of these
events, and the conclusion that the stellar obliquity is low.  The
short-cadence data also allow for better resolution of the
ingress/egress phases of the transit, leading to improved estimates of
the basic transit parameters.  The data can also be searched for
occultations, which would reveal the planet's dayside brightness; and
for ellipsoidal variations (ELV) and the effects of Doppler boosting
(DB), the amplitudes of which can be used to make independent
estimates of the planetary mass.  Finally, the continuous sequence of
transit times permits a search for any variations in the intervals
between transits, which could be caused by additional orbiting bodies
or tidal effects.

The paper is organized in the following way.
Section \ref{sec: photo} describes our reduction of {\it K2} data.
Section \ref{sec: refine} lays out the analysis of the light curve and the refinement of transit parameters.
Section \ref{sec: ttv} presents a search for changes in the transit period.
Section \ref{sec: rot} discusses the measurement of the stellar
rotation period, and the associated ``gyrochronological'' age.
Section \ref{sec: elv} presents the search for occultations, ELV, and DB effects.
Section \ref{sec: anomalies} presents the
analysis of spot-crossing anomalies and the implications for the stellar obliquity.
Finally, Section \ref{sec: dis} summarizes and discusses all our
findings.

While this work was in the final stages of preparation, we became
aware of the work of \citet{Mocnik}, who performed a similar analysis
of the same data.  Our study and their study have reached similar
conclusions regarding the stellar obliquity, stellar rotation period,
transit-timing results, and flux modulation outside of transits.  Some
small differences exist in the quantitative results, which we describe
in the appropriate sections.

\section{K2 Photometry}
\label{sec: photo}

\begin{figure*}
\begin{center}
\includegraphics[width = 2.2\columnwidth]{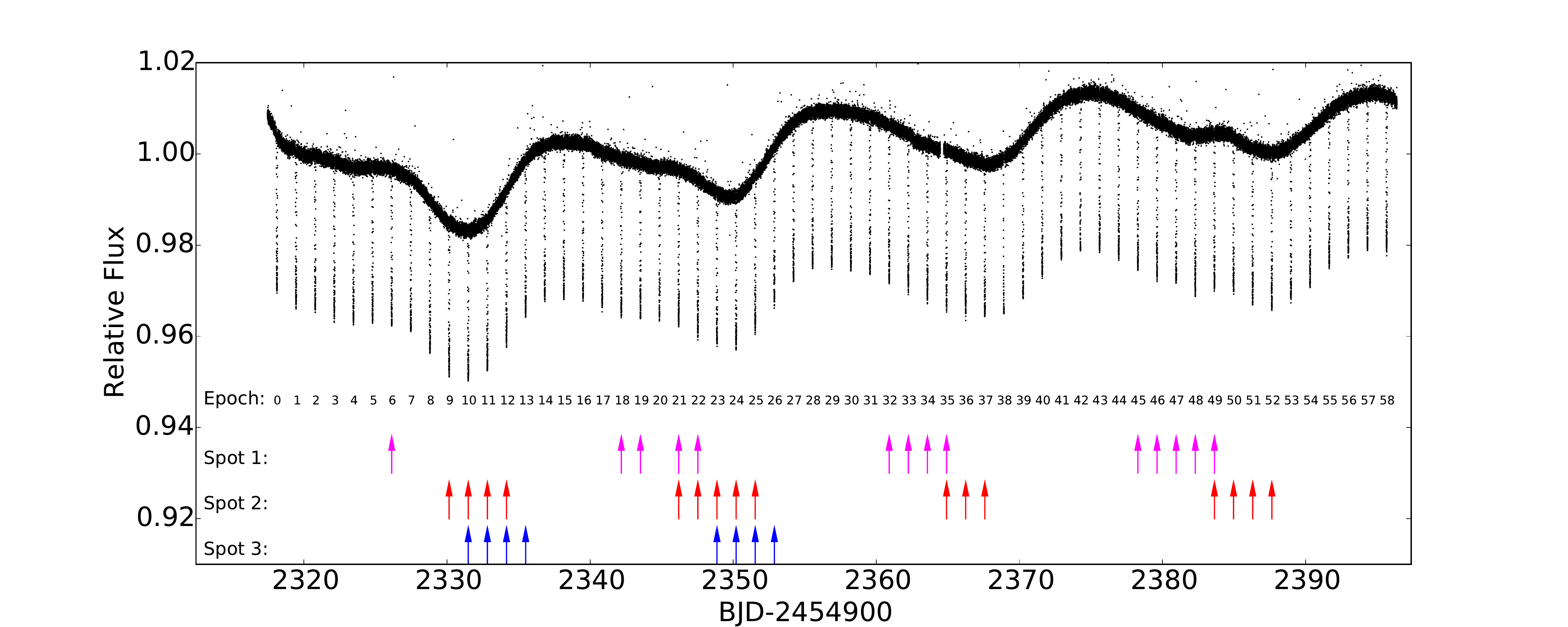}
\caption{
  Corrected {\it K2} photometry of Qatar-2.  Colored arrows indicate
  the times of identified spot-crossing anomalies (see
  Table \ref{tab: anomalies} for the full list of anomalies and Section \ref{sec: geometric}
  for how the anomalies were grouped).
  Anomalies recur in groups of $\approx$4, on a timescale similar to the stellar
  rotation period of $\approx$18 days.
}
\label{fig: lc}
\end{center}
\end{figure*}

Qatar-2 (or EPIC~212756297) was observed during {\it K2} Campaign 6
from 2015~Jul~11 to Oct~3 in the short-cadence mode. We downloaded the
pixel files from Mikulski Archive for Space Telescopes (MAST) website.
As is now well known, the photometric precision of {\it K2} data is
not as good as the original {\it Kepler} mission, due to the
uncontrolled rolling motion around the telescope's boresight combined
with the inter-pixel and intra-pixel sensitivity variations
\citep{Howell2014}.  To produce a photometric time series from the
pixel-level data, we used an approach similar to that described by
\citet{VJ2014}. In short, we used a circular aperture of 4.5 pixels in
radius centered around the brightest pixel.  To determine the
flux-weighted center of light, we fitted a two-dimensional Gaussian
function to the flux distribution of the pixels within this aperture.
We then fitted a piecewise linear function between the aperture-summed
flux and the coordinates of the center of light, and used the
parameters of the best-fitting function to correct the aperture-summed
flux time series.  Fig.~\ref{fig: lc} shows the corrected time series.

\section{Refining transit parameters}
\label{sec: refine}

The high precision and high temporal sampling rate provided by {\it
  K2} short-cadence observations are ideal for resolving the ingress
and egress phases of the transit, as well as revealing any anomalies
in the transit profile. Before searching for anomalies, we used the
short-cadence light curve to refine the basic transit parameters of
Qatar-2b. Because the extant radial-velocity data are consistent with
a circular orbit \citep{Bryan2012}, we assumed the orbit to be circular orbit in all our analyses.\footnote{The orbital
  eccentricity can also be constrained from the timing of the
  secondary eclipse; however, we did not detect the signal of the
  secondary eclipse in the {\it K2} data (see Section \ref{sec:
    elv}).}

We started with the corrected {\it K2} light curve (Fig.~\ref{fig:
  lc}) and published transit parameters \citep{Bryan2012}. We first
analyzed each transit individually by isolating a 7-hour window around
the expected mid-transit times. To remove the long-term stellar
variability, we allowed the flux baseline to be a quadratic function
of time, in addition to modeling the loss of light due to the
planetary transits.  For the transit model, we used the {\tt Python}
package {\tt Batman} by \citet{Kreidberg2015}. We adopted a quadratic
limb-darkening profile. We chose not to impose any priors on the two
limb-darkening coefficients because the short-cadence data proved to
provide adequate constraints on both coefficients (see Table~\ref{tab:
  para}).

Another effect that alters the transit profile is the presence of
starspots outside of the transit chord. Transit models such as {\tt
  Batman} assume the photosphere to be unspotted. When spots are
present, the untransited portion of the photosphere makes a smaller
relative contribution to the total flux than is assumed in the
model. If this is not accounted for, the model parameters would
compensate for the relatively large loss of light by increasing the
planet size, giving a biased result. To account for this effect, we
introduced an additional parameter specific to each transit:
$\Delta F_{\text{spot}}$, the relative loss of light due to any unocculted
spots on the visible hemisphere.
The calculated flux that is compared to the observed flux is
\begin{equation}
  F_{\text{calc,~spot}} = \frac{F_{\text{calc,~no-spot}} - \Delta F_{\text{spot}}}
  {1 - \Delta F_{\text{spot}}}
\end{equation}
where $F_{\text{calc,~spot}}$  and $F_{\text{calc,~no-spot}}$ are respectively the theoretical flux with and without taking the unocculted starspots into account. In this equation, the role of the denominator is to ensure that
$F_{\text{calc,~spot}} \equiv 1$ outside of the transits, since the data
have been normalized in this manner.

In summary, the set of parameters describing each transit are the time
of inferior conjunction ($T_{\text{c}}$); the three parameters of the
quadratic function of time representing stellar variability ($a_2$,
$a_1$, and $a_0$); and the loss of light due to unocculted spots on
the visible hemisphere ($\Delta F_{\text{spot}}$). There are also the
usual transit parameters: the planet-to-star radius ratio
($R_p/R_\star$); the ratio of stellar radius to orbital distance
($R_\star/a$); the impact parameter ($b$), and the limb-darkening
coefficients ($u_1$ and $u_2$). We adopted the usual $\chi^2$
likelihood function and found the maximum-likelihood solution using
the Levenberg-Marquardt algorithm as implemented in the {\tt Python}
package {\tt lmfit} \citep{lmfit}.

\begin{figure*}
\begin{center}
\includegraphics[width = 2.2\columnwidth]{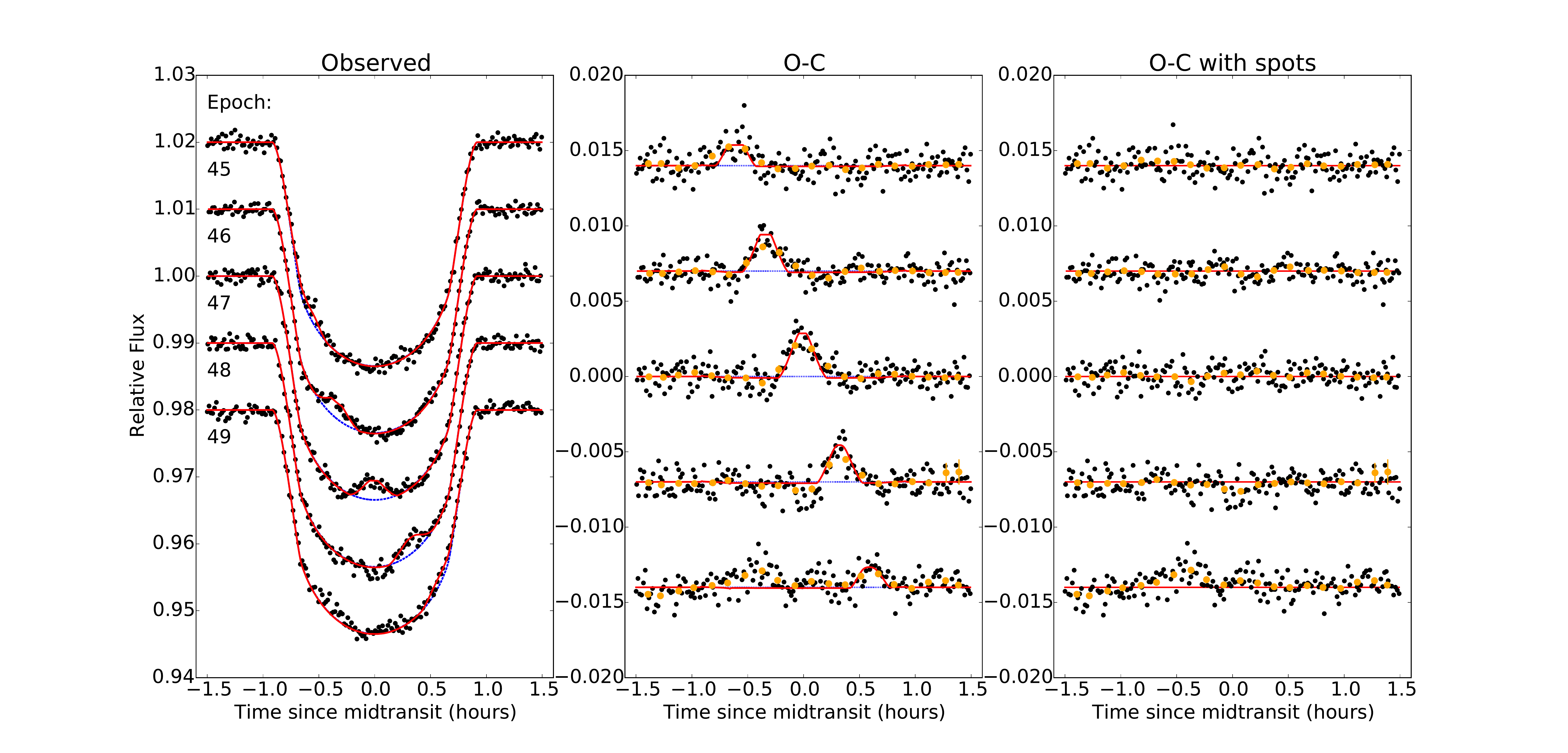}
\caption{
  {\it Left.}---The five consecutive transit light curves that were selected
  for detailed modeling.
  Arbitrary vertical offsets have been applied to data from different epochs.
  The dotted blue line is the best-fitting model with no spots.
  The solid red line is the best-fitting single-spot model.
  {\it Middle.}---Residuals, after subtracting the no-spot model.
  The spot-crossing anomalies are seen to progress steadily in phase
  from one transit to the next, at the rate that is expected, based on the
  orbital period and the stellar rotation period.
  {\it Right.}---Residuals, after subtracting the single-spot model.
}
\label{fig: combined}
\end{center}
\end{figure*}

Spot-crossing anomalies are clearly visible in the time series of
residual fluxes. Fig.~\ref{fig: combined} shows some examples. These anomalies
would be a source of bias in the model parameters, if no corrections
were performed.
We identified these anomalies through visual inspection, and modeled
them as Gaussian functions of time:
\begin{equation}
F_{\text{anom}}(t) = A \exp{\left[-\frac{(t-t_{\text{anom}})^2}{2\sigma_{\text{anom}}^2}\right]}
\end{equation}
where $A$, $t_{\text{anom}}$ and $\sigma_{\text{anom}}$ represent (respectively) the
amplitude, time, and duration of the anomaly.

In some cases, visual inspection of a given transit revealed more than
one spot-crossing anomaly. To decide on the number of spot-crossing
anomalies to include in the final model, we fitted the
light curve with increasing numbers of spots, and calculated the
change in the Bayesian Information Criterion,
\begin{equation}
\text{BIC}  = 2 \text{log}(L_{\text{max}})+N~\text{log}(M),
\end{equation}
where $L_{\text{max}}$ is the maximum likelihood, $N$ is the number of
model parameters, and $M$ is the number of data points.  We only
retained those anomalies for which $\Delta$BIC~$>10$.  Table~\ref{tab:
  anomalies} reports the properties of all these anomalies.  The
parameter uncertainties were determined via the Markov Chain Monte
Carlo (MCMC) method, as implemented in the {\tt Python} package {\tt
  emcee} \citep{emcee}. Here and elsewhere in this paper, the reported
parameter value is based on the 50\% level of the cumulative posterior
distribution, and the uncertainty interval is based on the 16\% and
84\% levels.

\begin{figure}
\begin{center}
\includegraphics[width = 1.\columnwidth]{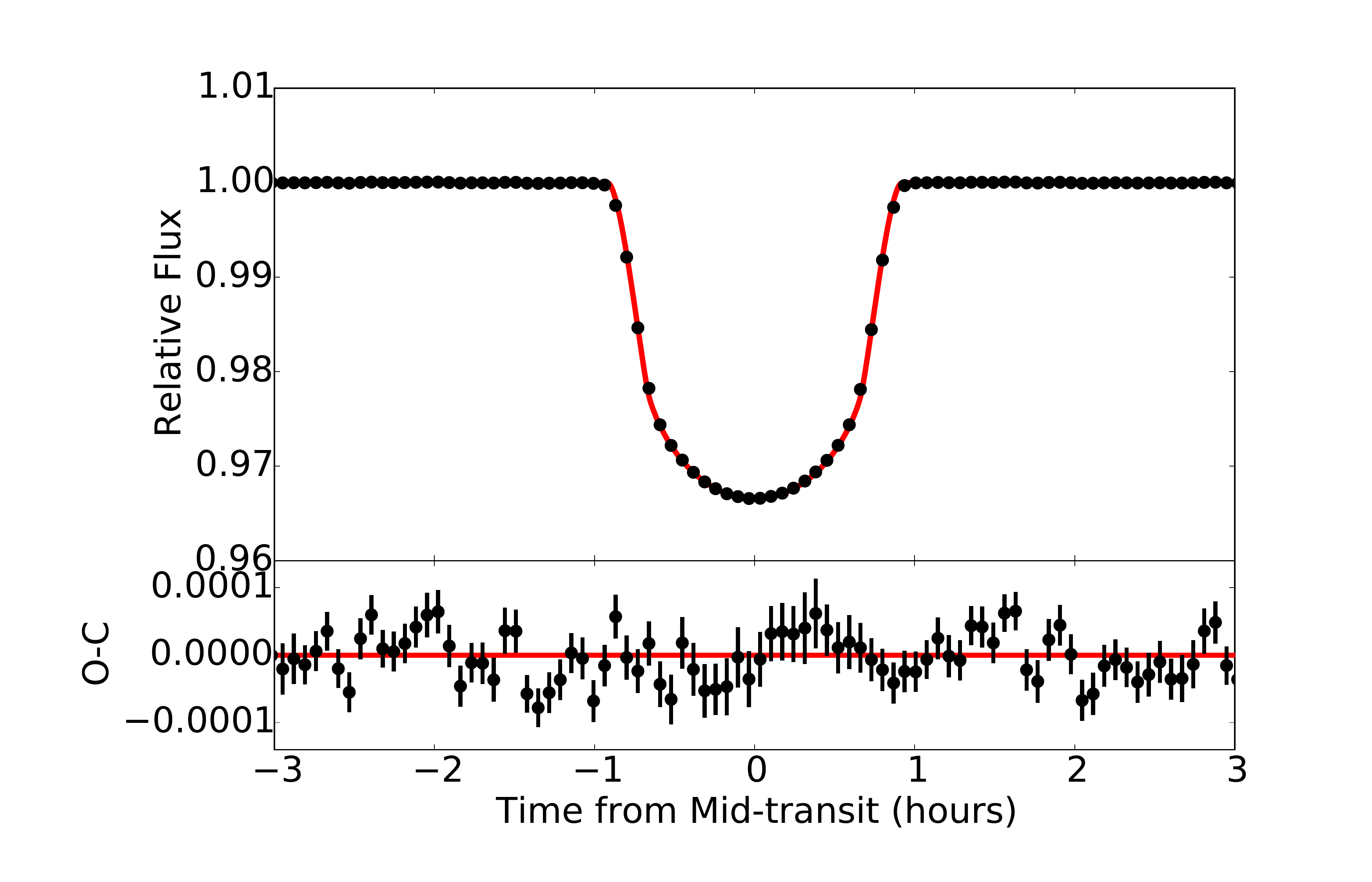}
\caption{{\it Top.}---Phase-folded transit light curve of Qatar-2, after correcting
  for stellar variability, unocculted spots, and spot-crossing anomalies.
  The red line shows the best-fitting model. The data have been averaged in phase
  within intervals of 3 minutes. 
  {\it Bottom.}---Residuals between the data and the best-fitting model.}
\label{all}
\end{center}
\end{figure}

We used the best-fitting parameters to correct the data from each
transit for stellar variability and unocculted spots.  We also removed
the spot-crossing anomalies by excluding data points within
$2\sigma_{\text{anom}}$ of the time of each anomaly.  We combined
all 59 of the rectified and spot-cleaned transit intervals to create a
phase-folded transit light curve with a very high signal-to-noise
ratio.  Then we modeled this phase-folded light curve to determine
the basic transit parameters, using another MCMC analysis ( see
  Fig.~\ref{all}).

We then assumed that these basic transit parameters are fixed in time
and applicable to each and every transit.  We repeated the analyses of
all of the individual transits, holding the transit parameters fixed
at the values determined from the analysis of the phase-folded light
curve.  This in turn allowed the creation of a new version of the
phase-folded light curve.  After two such iterations it was clear that
the results had already converged.  Table~\ref{tab: para} gives the
results.

\section{Lack of Transit Timing Variations}
\label{sec: ttv}

To search for evidence of any changes in the orbital period since the
time of discovery of Qatar-2b, we combined our measured midtransit times with
those found on the Exoplanet Transit Database (ETD) website.  Table
\ref{tab: ttv} gives all the midtransit times in the Barycentric
Dynamical Time system \citep[$\text{BJD}_{\text{TBD}}$]{Eastman2010}.

\begin{figure}
\begin{center}
\includegraphics[width = 1.\columnwidth]{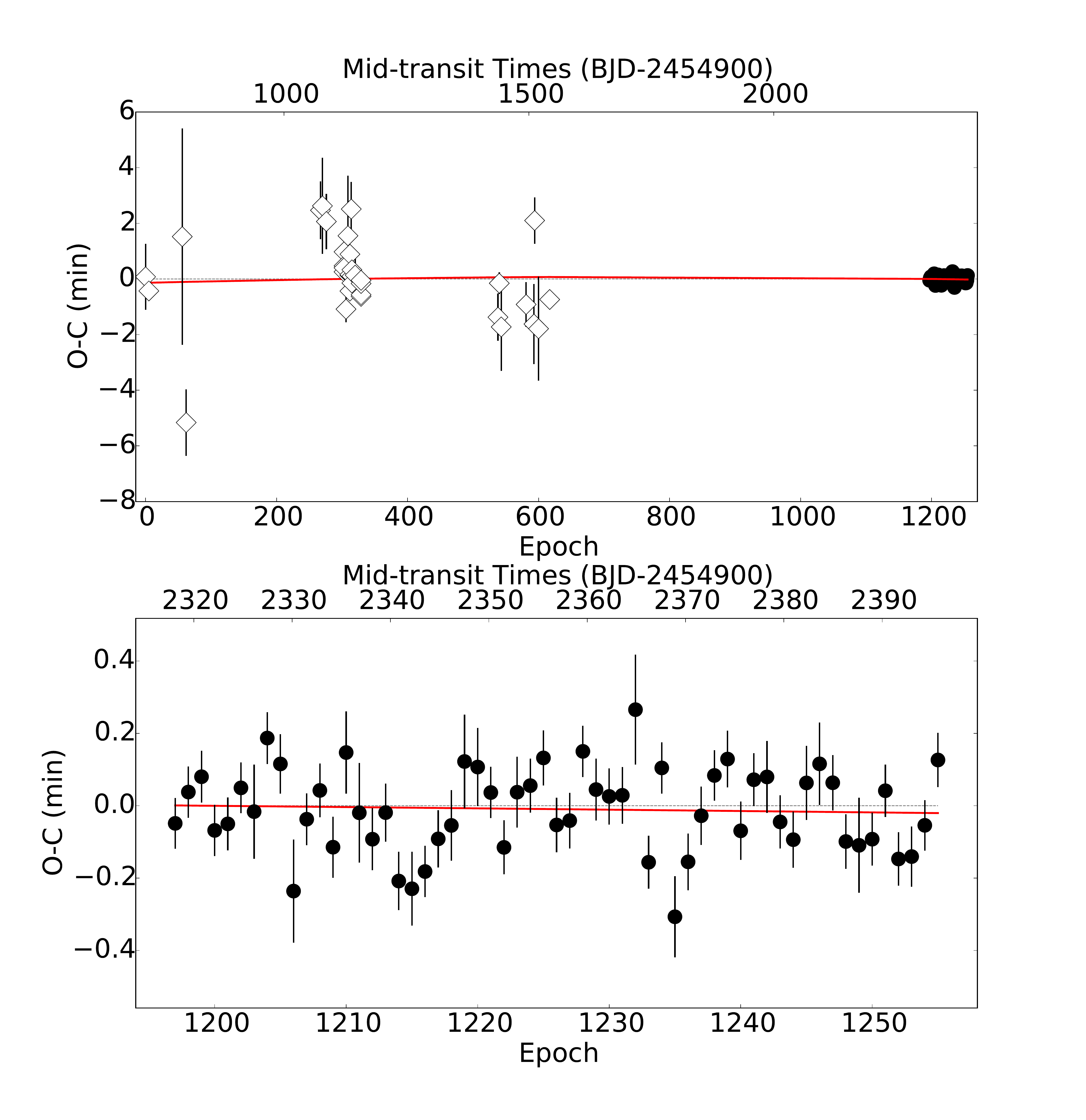}
\caption{
  Observed minus calculated transit times of Qatar-2b, where the
  calculated times are based on the best-fitting constant-period model.
  The top panel shows all the available data, and the bottom panel
  allows a closer view of the {\it K2} data.
  Table \ref{tab: ttv} gives the numerical data.
}
\label{fig: ttv}
\end{center}
\end{figure}

Fig.~\ref{fig: ttv} shows the residuals between the observed times and
the calculated times according to the best-fitting constant-period
model.  The only obvious pattern in the residuals is that the data
points from the second season are generally above the baseline, while
the third season's data are below the baseline. It will be interesting
to see if these long-term variations are seen in future seasons.  We
do not find any sinusoidal-like variations that are sometimes seen in
multi-planet systems.  We computed the Lomb-Scargle periodogram
  \citep{Lomb1976,Scargle1982} of the timing residuals; no signal was
  detected with a false alarm probability less than 10\%.  We
  also did not detect any evidence for a secular change in the orbital
  period, as described below. The lack of detectable period shrinkage
  allows us to place a constraint on the rate of tidal dissipation in
  the system.  Tidal evolution is expected to cause period decay with
a rate that scales as $(M_{\text{p}}/M_\star)(R_\star/a)^{5}$
\citep{Levrard2009}, which is relatively large for this system because
of the close-in orbit. For quantitative constraints on the rate, we
fitted the following function to the sequence $T_n$ of midtransit
times:
\begin{equation}
\label{eqn: quad}
T_n = T_0 + n P_0 + \frac{1}{2} n^2 \frac{dP}{dn}.
\end{equation}
We conducted a MCMC analysis using {\tt
  emcee} and the usual $\chi^2$ likelihood function, and uniform priors for
all parameters.  The result for the period-change parameter
was an upper limit,
$|\frac{dP}{dN}|  < 0.11$~milliseconds, or
$|\frac{dP}{dt}|  < 1.7 \times 10^{-9}$ (95\% conf.).
To translate these upper bounds into a lower bound on the
the stellar tidal quality factor we used the formula \citep{Levrard2009} 
\begin{equation}
  Q'_\star = 9P^2 \left(\frac{dP}{dN}\right)^{-1}
  \frac{M_p}{M_{\star}}\left(\frac{R_{\star}}{a}\right)^{5}
  \left(\omega_\star-\frac{2 \pi}{P}\right),
\end{equation}
where $\omega_\star$ is the angular velocity of stellar rotation.  The
derivation of this formula assumes a circular orbit and zero
obliquity.  For Qatar-2, a low eccentricity is compatible with the
available radial velocity dataset \citep{Bryan2012}, and a low
obliquity is implied by our analysis in Section \ref{sec: anomalies}.
The result of applying this formula to our data is $Q_\star^{'}~>1.5
\times 10^4$ (95\% conf.).

\section{Stellar rotation period and gyrochronology}
\label{sec: rot}

\begin{figure}
\begin{center}
\includegraphics[width = 1.\columnwidth]{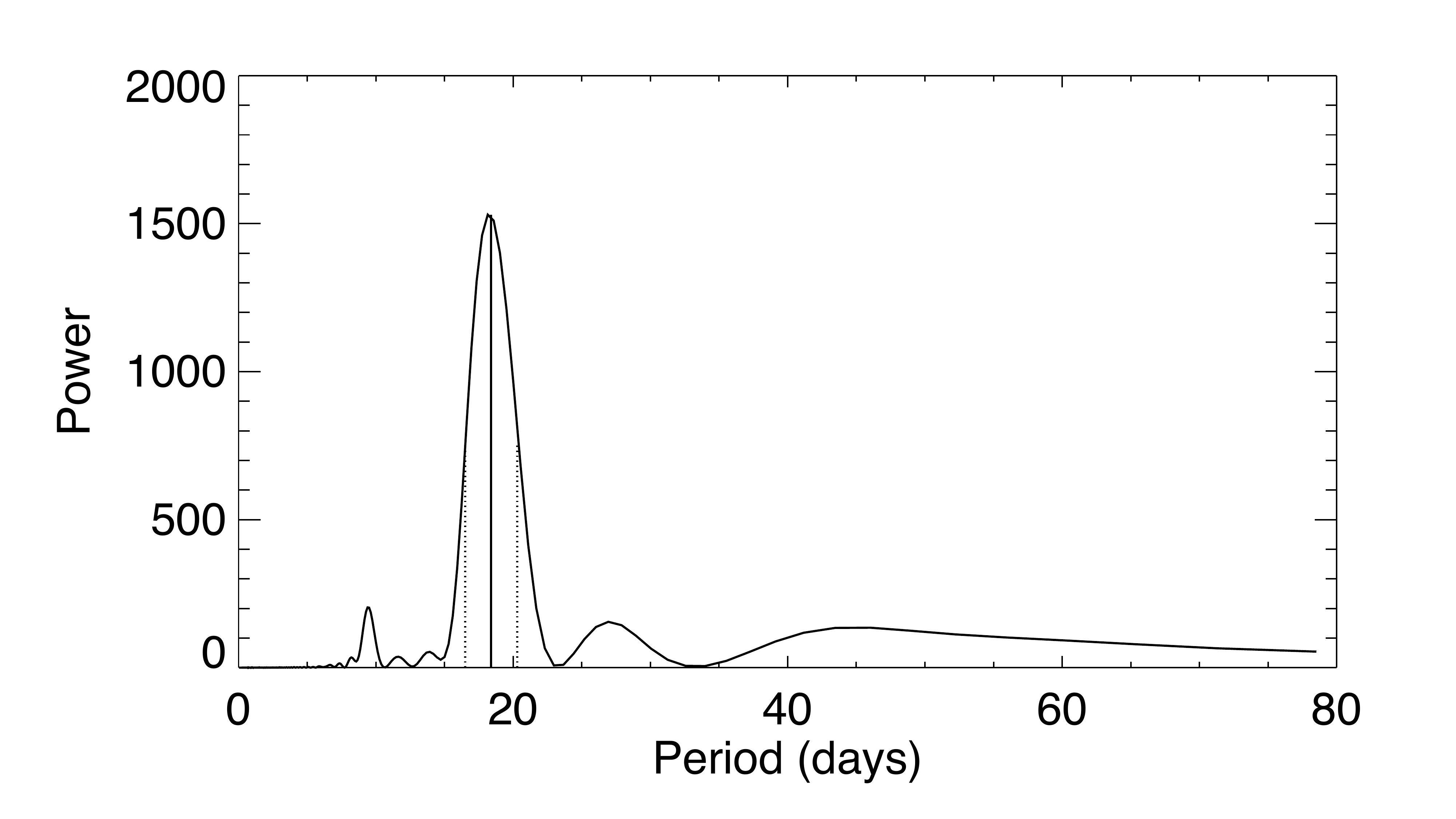}
\caption{Lomb-Scargle peridogram of the out-of-transit flux
  variations of Qatar-2.  Based on the location and width of the most
  prominent peak we estimate the rotation period to be $18.5 \pm
  1.9$~days.}
\label{fig: per}
\end{center}
\end{figure}

The {\it K2} light curve (Fig.~\ref{fig: lc}) exhibits quasiperiodic flux
variations with four cycles. These variations are characteristic of
starspots being carried around by rotation, and therefore the stellar
rotation period can be estimated from the period of these flux
variations. For a quantitative estimate, we masked out the transits
and calculated the Lomb-Scargle periodogram
\citep{Lomb1976,Scargle1982} of the resulting time series, which is
shown in Fig.~\ref{fig: per}. Based on the location and width of the
most prominent peak in the periodogram, we estimate the stellar
rotation period to be $18.5 \pm 1.9$~days.

Knowledge of the stellar rotation period played a crucial role in our
obliquity determination (see Section \ref{sec: anomalies}).  In
addition, for main-sequence stars such as Qatar-2, the rotation period
is linked to the stellar age, a relationship that has come to be known
as "gyrochronology.'' We estimated the age of the system using a
gyrochronological formula that was derived by \citet{Schlaufman2010}:
\begin{equation}
P_\star(M_\star, \tau_\star) = P_{\star,0}(M_\star)\left(\frac{\tau_\star}{650~\text{Myr}}\right)^{1/2},
\end{equation}
where $P_\star(M_\star, \tau_\star) $ is the rotation period of a star with mass
$M_\star$ and age $\tau_\star$, and $P_{\star,0}(M_\star)$ is a specified polynomial
function that was calibrated using data from the Hyades and
Praesepe star clusters. Using this formula and our measured rotation period,
the gyrochronological age of Qatar-2 is $1.4 \pm 0.3$~Gyr.

\citet{Maxted2015} made an independent estimate of the stellar age by
fitting stellar-evolutionary models to the observed spectroscopic
parameters and apparent magnitudes. Their result was $15.7 \pm
1.4$~Gyr, significantly older than the gyro age.  Assuming this older
age is correct, the younger gyro age could be taken as evidence that
the star has been spun up by the tidal torque of the close-in
planet. However, \citet{Maxted2015} expressed concern that their
estimate is unrealistic because their method may be biased by the
``inflated K-dwarf'' phenomenon, a known problem with
stellar-evolutionary models in fitting the observed properties of
stars similar to Qatar-2.

\citet{Mocnik} also used {\it K2} data to determine the stellar
rotation period, and found the gyro age to be $0.59\pm 0.10$~Gyr. This
is significantly younger than our estimate of the gyro age.  Since
their result for the rotation period was essentially equivalent to
ours, the difference in gyro ages must be attributable to the
different gyrochronological formula that was adopted by
\citet{Mocnik}.  They used a formula presented by \citet{Barnes2007},
while we used the formula above from \citet{Schlaufman2010}. Evidently
the gyro age is subject to a systematic uncertainty that is more
important than the uncertainty in the stellar rotation period.

Another use for the stellar rotation period is to estimate the
inclination $i_\star$ between the stellar rotation axis and the line
of sight. This is done through the following formula:
\begin{equation}
  \sin i_\star = \frac{v\sin i_\star} {v} = \frac{v\sin i_\star}{2\pi R_\star/P_{\text{rot}}},
\end{equation}
where $v\sin i_\star$ is the projected rotation rate that can be
estimated from the degree of rotational broadening that is observed in
the star's photospheric absorption lines.  For Qatar-2,
\citet{Bryan2012} found $v\sin i_\star = 2.8 \pm 0.5$~km~s$^{-1}$,
while our results lead to $v = 2\pi R\star / P_{\rm rot} = 2.0 \pm 0.3$~km
s$^{-1}$, giving $\sin i_\star = 1.4 \pm 0.6$. This is compatible with
unity, as expected for a low-obliquity star, although the uncertainty
is large enough to encompass inclinations as low as $50^\circ$ (as
well as mathematically impossible values of $\sin i_\star$).

\section{Phase curve analysis and secondary eclipse}
\label{sec: elv}

Thanks to the high precision and nearly continuous temporal coverage
of the {\it K2} data, we may perform a sensitive search for the
occultation of Qatar-2b by its parent star (the secondary eclipse), as
well as out-of-eclipse light variations associated with the orbital
period.  The latter type of variations could arise from the
tidally-induced ellipsoidal figure of the star (ELV), Doppler boosting
(DB), and illumination effects (ILL), as exemplified by
\citet{Mazeh2010}.  All of these effects are expected to be small, on
the order of $\sim$10~ppm.  Thus it is difficult, and important, to
try and distinguish any residual systematic effects in the time series
from the astrophysical effects.

For this reason we performed all our analyses on several versions of
the {\it K2} light curve, all of which used different algorithms to
try and filter out systematic effects and artifacts.  Specifically we
used the versions known as {\tt K2SFF} \citep{VJ2014, Vanderburg2016},
{\tt K2SC} \citep{Aigrain2016, Pope2016}, {\tt K2 Everest}
\citep{Luger2016}, and our own processed light curve. We found that
while all of the light curves gave consistent results, {\tt K2
  Everest} seemed to have the lowest levels of residual systematic
trends and artifacts. This particular algorithm differs from all the
others by being based on Pixel Level Decorrelation
\citep[PLD]{Deming}.  All the other methods rely on measurements of
the flux-weighted center of light of a specificed collection of
pixels.  The results described in the rest of this section are based
on the {\tt K2 Everest} light curve.

\begin{figure}
\begin{center}
\includegraphics[width = 1.\columnwidth]{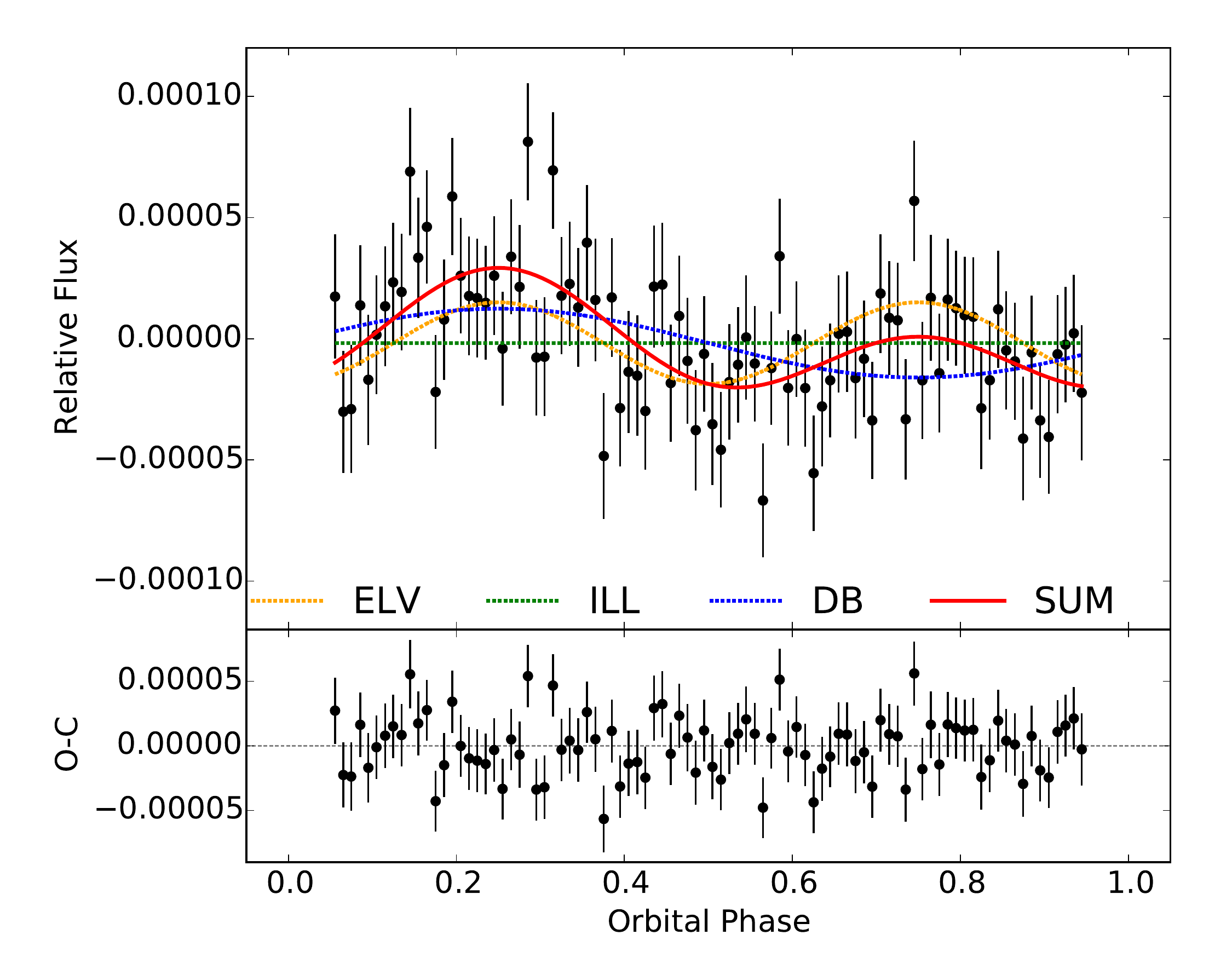}
\caption{ The phase-folded and binned light curve of Qatar-2, after
  excluding the transits and filtering as described in the text.  The
  red curve shows the best-fitting model. The different components of
  the model are shown in different colors (except the additive constant, which is not shown.)
  No secondary eclipse was
  detected.  }
\label{phase}
\end{center}
\end{figure}

We omitted all the data within 3~hours of each midtransit time.  To
remove the long-term stellar variability, we divided the light curve
by a cubic spline with a temporal width of twice the orbital
period. We then folded the time series with the orbital period of the
planet, and averaged the resulting light curve into 100 bins equally
spaced in orbital phase. We fitted for the ELV, DB and ILL effects
simultaneously  (See Fig.~\ref{phase}). For the ILL component, we adopted a Lambertian phase
function.  The combined model for the variations took the form
\begin{equation}
 F_0-A_{\text{ELV}}\text{cos}(4\pi\phi)+A_{\text{DB}} \text{sin}(2\pi\phi) + A_{\text{ILL}}\frac{\text{sin}(z)+(\pi-z)\text{cos}(z)}{\pi}
\end{equation}
where
\begin{equation}
\text{cos}(z) = -\text{sin}(i)\text{cos}(2\pi\phi+\theta)
\end{equation}
and
\begin{equation}
\phi = \frac{t-T_\text{c}}{P}.
\end{equation}
In these equations,  $F_0$ is an additive constant, 
$A_{\text{ELV}}$, $A_{\text{DB}}$ and $A_{\text{ILL}}$ are the amplitudes of the ELV, DB and ILL effects,
$T_\text{c}$ is the time of inferior conjunction, $P$ is the orbital period,
$i$ is the orbital inclination, and $\theta$ represents a hypothetical offset between the
maximum of the phase curve and the time of superior conjunction.
We also fitted for the loss of light during the secondary eclipse, using {\tt Batman},
and requiring the depth of the secondary eclipse to be consistent with $A_{\text{ILL}}$.
Initially, we allowed the phase of the secondary eclipse to be a free parameter; once it became
clear that no secondary eclipse could be detected, we reverted to the assumption
of a circular orbit and thereby required the secondary eclipse
to occur at $\phi=0.5$.

We conducted an MCMC analysis with {\tt emcee}, with uniform priors on
all the parameters. The $A_{\text{ELV}}$ and $A_{\text{DB}}$
parameters were both found to be nonzero. Specifically, $A_{\text{ELV}} = 15.4 \pm 4.8$~ppm and
$A_{\text{DB}} = 14.6 \pm 5.1$ ppm.  Both of these effects depend on
the planet mass, along with additional system parameters that are
largely constrained by other observations. Therefore we may use the
results for $A_{\text{ELV}}$ and $A_{\text{DB}}$ to make independent
determinations of the planet mass. For this purpose we used Equations
(11), (12) and (15) of \citet{Carter2011}. The mass implied by the ELV
amplitude is $M_{\text{p, ELV}} = 2.6 \pm 0.9 ~M_{\text{Jup}}$, while
the mass implied by the DB amplitude is less certain $M_{\text{p, DB}} = 3.9 \pm
1.5~ M_{\text{Jup}}$. These two independent estimates are consistent with each other
to within one sigma, and also agree with the mass determination $M_{\text{p,RV}} =
2.487 \pm 0.086 ~M_{\text{Jup}}$ based on the more secure and
traditional Doppler technique \citep{Bryan2012}.   This lends
confidence to our assessment that the out-of-transit flux variations
are astrophysical rather than being dominated by instrumental or
systematic effects.

Neither the ILL effect nor the secondary eclipse were detected. The
resulting upper bound on $A_{\text{ILL}}$ is 35~ppm (95\% conf.).
This represents an upper bound on the combination of the planet's
reflected light and thermal emission.  Assuming that the thermal
emission is negligible within the {\it Kepler} bandpass, the resulting
upper limit on the planet's geometric albedo is $A_g <   0.06$.  Any
contribution from thermal emission would require an even smaller
geometric albedo.  Conversely, if the reflected component is assumed
to be negligible we may place an upper bound on the effective
temperature of the planet, after making the simplifying assumption
that the planet emits as a blackbody.  The resulting upper limit is
$T_{\text{eff}} < 1500$~K (95\% conf.).  This is consistent with the
calculated equilibrium temperature of $T_{\text{eq}} \approx 1300~K$,
assuming a Bond albedo of zero.

\section{Spot-crossing Anomalies and obliquity measurement}
\label{sec: anomalies}

In this section, we present the analysis of the spot-crossing
anomalies.  The patterns of recurrence of the spot-crossing anomalies
implies that the transit chord is aligned with the lines of latitude
on the star, which in turn implies that the star has a low obliquity.
For quantitative analysis we employed two different approaches, each
of which has its advantages and limitations.

\subsection{Anomaly identification and timing}
\label {sec: geometric}

First we employed a simple geometric model for which the parameters
are constrained by the measured times of spot-crossing anomalies.
Similar models have previously been used to constrain the obliquity of
planet-hosting stars \citep{Sanchis-Ojeda2011Wasp, Nutzman2011}. The
premise is straightforward: when an anomaly is observed, the planet's
position on the sky must at least partially overlap the location of
the starspot. We define our coordinate system in the plane of sky such
that the $x$-axis is aligned with the line of nodes of the planetary
orbit, and the $y$-axis is in the perpendicular direction.  Using the
basic transit parameters determined earlier, we calculate the
projected $x$ and $y$ coordinates of the planet as a function of time.
We choose a particular spot-crossing anomaly as the nominal starting
point, at which the starspot is placed at the position of the
planet. Then we can predict any future or past location of the
starspot, given the following parameters: the stellar inclination
($i_{\star}$), the sky-projected obliquity ($\lambda$), the stellar
rotation period ($P_{\text{rot}}$), and the stellar latitude of the
spot ($l$).  For simplicity we assume that the starspot does
not change significantly in size, intensity, or location during the
interval over which the model is applied. This assumption is more
valid when focusing on a relatively short time interval.

\begin{figure}
\begin{center}
\includegraphics[width = 1.\columnwidth]{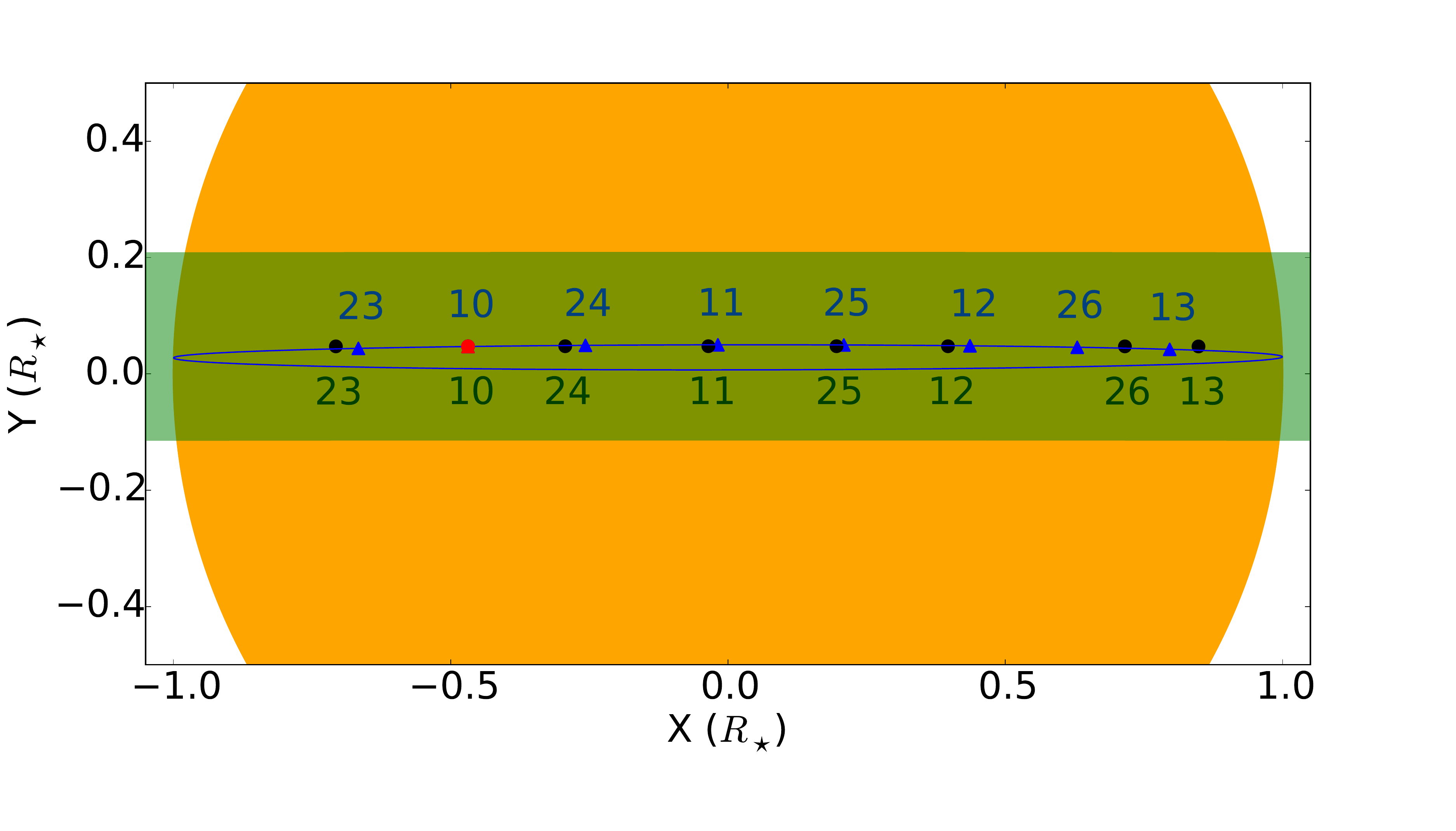}
\caption{ Illustration of the first modeling approach, in which we
  measure the times of anomalies that seem associated with a single
  spot, and constrain a geometric model by requiring the planet and
  spot to be nearly coincident on the sky plane at those times.  The
  orange circle represents the stellar disk.  The green band is the
  transit chord.  The black circles are the calculated positions of
  the planet during each spot-crossing anomaly that was assigned to
  Spot 3 from Table~\ref{tab: anomalies}.  The numbers specify the
  transit epoch numbers.  The blue curve is the calculated trajectory
  of Spot 3, and the blue triangles are the calculated locations of
  Spot 3 at the time of the anomalies.  The success of the model is
  illustrated by the near-coincidence of the planet and spot at the
  times of all the anomalies.  }
\label{fig: geometric}
\end{center}
\end{figure}

Fig.~\ref{fig: geometric} illustrates this model, using the anomalies
associated with Spot 3 in Table \ref{tab: anomalies}.  The black dots
show the calculated positions of the planet during each spot-crossing
anomaly caused by Spot 3.  We initialized the model by assuming that
the spot and planet coincided at the time of the first anomaly (Epoch
10, red circle). The blue curve shows the spot's trajectory on the
stellar photosphere, and the blue triangles show the calculated
positions of the spot at the times of the observed anomalies.  The
success of the model is indicated by the close coincidence between the
positions of the planet and spot.

A key question is what to do when there are multiple spots on the
star, which is likely in general, and definitely the case for Qatar-2.
When the model has multiple spots, how do we associate individual
spot-crossing anomalies with a particular spot?  First we grouped the
spot-crossing anomalies into families through visual inspection of
their relative phases, amplitudes and durations.  We then revised
these assignments as needed when the model revealed significant
outliers indicating a mistaken association.  Our final assignments are
justified by noting that the analysis of all the different groups were
consistent with the same rotation period, which is in turn consistent
with the rotation period estimated from the {\it K2} light curve (see
Table \ref{tab: geometric}).

Although this procedure seems to work, the necessity to group the
anomalies as we have just described is a shortcoming of this simple
geometric model.  This weakness is especially acute when the technique
is applied to ground-based data, for which quasi-continuous monitoring
is very difficult to achieve.  As an example, \citet{Mancini2014} did
not have the stellar rotation period as an independent check for their
model of Qatar-2b. By assuming that two particular spot-crossing
anomalies they observed were associated with a single spot, they
derived a stellar rotation period of $14.8 \pm0.3$ days
  \citep[as later revised by] []{Mancini2016}. This is now known to be
incorrect; most likely, the two observed anomalies were produced by
crossings over two different spots.

For quantitative constraints on the obliquity,
we adopted the likelihood function
\begin{equation}
L = \exp{(-\chi^2/2)},
\end{equation}
where
\begin{equation}
\label{eqn: like}
\chi^2 = \sum_{i}^{N_{\text{anom}}}{\frac
  {(x_{\text{spot,i}}-x_{\text{p,i}})^2+(y_{\text{spot,i}}-y_{\text{p,i}})^2}
  {(0.5~R_p)^2}} + {\rm NDP}.
\end{equation}
Here, $N_{\text{anom}}$ is the number of spot-crossing anomalies; and
$x_{\text{spot,i}}$, $y_{\text{spot,i}}$, $x_{\text{planet,i}}$ and
$y_{\text{planet,i}}$ are the coordinates of the spot and the planet
at the time of the $i$th anomaly.  With this function, we reward
models that place the planet and spot close to one another at the
times of observed anomalies.  By choosing the length scale to be
$0.5~R_p$ we have assumed that the spot sizes are comparable to the
size of the planet, or smaller. The NDP term is the non-detection
penalty, which adds 100 to $\chi^2$ if there is no observed anomaly at
a time when the model predicts one. We acknowledge that the choice of
length scale and NDP are {\it ad hoc}, preventing the quantitative
results from being taken too seriously; the purpose of the modeling is
simply to demonstrate that low-obliquity solutions are able to account
for the most prominent sequences of anomalies.

To begin, we identified the three most prominent series of
spot-crossing anomalies (labeled with red, blue, and magenta arrows in
Fig.~\ref{fig: lc}), and analyzed each one of these families
separately with a one-spot model. Then after being satisfied that they
gave consistent results, we performed a joint analysis of all the
spot-crossing anomalies using a three-spot model. Table \ref{tab:
  geometric} gives all the results, based on an MCMC analysis.  We
reiterate that the quantitative results are contingent on the choices
of length scale and NDP in the likelihood function, which were chosen
somewhat subjectively. The main point is that in all cases, the
sky-projected obliquity is consistent with zero, and the stellar
rotation period is consistent with the independently measured period
of $18.5 \pm 1.9$~days.  The stellar inclination $i_{\star}$ and spot
latitude $l$ are only loosely constrained, and their uncertainties are
strongly correlated, demonstrating another limitation of this modeling
approach.

\subsection{Light-curve fitting}

\begin{figure}
\begin{center}
\includegraphics[width = 1.\columnwidth]{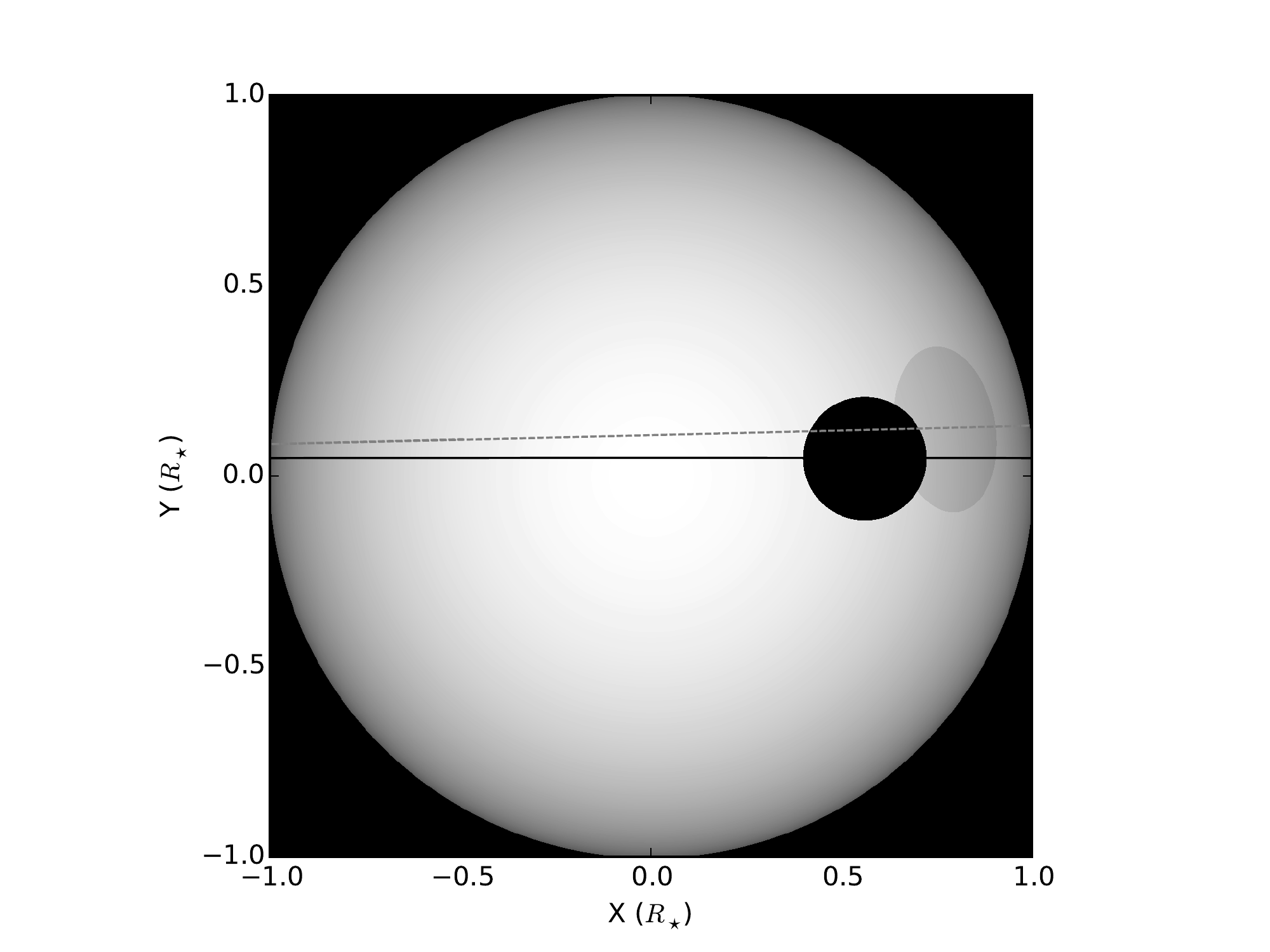}
\caption{ Illustration of the best-fitting spot model from
  Fig.~\ref{fig: combined}, at the time of Epoch 49. Visible are the
  limb-darkened photosphere, the starspot (light gray ellipse near the
  limb), and the planet (black circle).  The dashed line is the spot's
  trajectory across the stellar disk, and the solid line is the
  planet's trajectory.  A given spot produces a smaller anomalies when
  it is projected near the limb, due to geometrical foreshortening and
  limb-darkening.}
\label{fig: sky}
\end{center}
\end{figure}

As a second approach to demonstrate the low obliquity of Qatar-2, we
constructed a numerical model for the loss of light due to a planetary
transit over a star with circular starspots. We used a two-dimensional
Cartesian grid to represent the stellar disk, and assigned intensities
to the pixels based on the assumed limb-darkening law and the
locations of starspots and the planet (see Fig. \ref{fig: sky}).  For
simplicity the spots were assumed to be circular and uniform in
intensity, with unchanging properties and locations in the rotating
frame of the star. Thus, in addition to the usual transit parameters,
this model has parameters for the spot's angular size ($\alpha$),
intensity contrast ($f$), latitude ($l$), and the time ($t$) when it
crosses the $x$-axis.  We also allow each spot to be associated with
an independent rotation period, $P_{\text{rot,i}}$, to allow for a
consistency check (and to allow for a modest degree of differential
rotation, although we did not end up finding evidence for this
effect).

At any particular time, we located the pixels affected by the spot by
taking the dot product between the surface normal associated with the
pixel and the position vector of the spot. The intensity of any pixel
within the angular radius of a spot center was multiplied by the
spot's contrast factor. The pixels within the planet's silhouette were
assigned zero intensity.  Then the summed intensity of all the pixels
was compared to the observed flux, and the usual $\chi^2$ statistic
was calculated.  We held fixed the transit parameters at the
best-fitting values obtained in Section~\ref{sec: refine}.

The pixelated model is conceptually straightforward, but it requires
two-dimensional integration, which is computationally
expensive. \citet{Beky2014} wrote a semi-analytic code called {\tt
  Spotrod} to model spot-crossing anomalies, also assuming uniform and
circular spots. Their algorithm is more computationally efficient
because the integration is reduced to one dimension through the
analytic calculation of the points of intersection between the spot
and planet.  We analyzed the light curve with our own 2-d model as
well as {\tt Spotrod}, to check for consistency.

Although more than half of the transits observed by {\it K2} showed
evidence for spot-crossing anomalies, we chose to model five
consecutive anomalies with the highest signal-to-noise ratio.  We
chose to limit the time interval of the model to $\approx$5~days
because we are not modeling spot evolution.

First we found the maximum likelihood model, using the
Levenberg-Marquardt algorithm as implemented in {\tt lmfit}. We then
conducted a MCMC analysis with {\tt emcee}. Table \ref{tab: numerical}
gives the results.  The results from our 2-d numerical model and {\tt
  Spotrod} are very similar.  The stellar inclination was found to be
within about $10^\circ$ of edge-on, and the sky-projected obliquity
was found to be consistent with zero within about 5$^\circ$.  The
stellar rotation period was found to agree well with the value
reported previously in Section~\ref{sec: rot}. The angular radius of
the spot was around $10^{\circ}$, much larger than sunspots.

The numerical light-curve modeling may appear to offer very precise
constraints on the obliquity and other system parameters. However,
just as was the case with our first modeling approach, the precise
quantitative results should not be taken too seriously.  In this case
it is because the light curve models make strong assumptions about the
shape and intensity distribution of the spots, as well as the lack of
any spot migration or evolution.  There is no reason to believe that
the spots are circular, and indeed each "spot'' may in reality be a
complex, splotchy arrangement of spots and plages.  We regard the
numerical results as a conceptually straightforward demonstration that
the obliquity is likely to be smaller than about 10$^\circ$.

\section{Summary and Discussion}
\label{sec: dis}

In this work, we presented the analysis of the {\it K2} short-cadence
observation of Qatar-2. The continuous monitoring, high precision and
high cadence of the {\it K2} data helped to refine the transit
parameters. In addition, the data quality was high enough to
facilitate the identification and exclusion of data points affected by
spot-crossing anomalies, leading to a less biased set of transit
parameters.

We measured the stellar rotation period of Qatar-2A $18.5 \pm 1.9$
days based on the out-of-transit flux variation of the {\it K2} light
curve.  Using the technique of gyrochronology, the rotation period led
to an independent estimate of the stellar age, $1.4 \pm 0.3$ Gyr. The
rotation period also played a crucial role in our obliquity
determination; the lack of an independently measured rotation period
had been a missing piece of the puzzle in a previous effort to
determine the stellar obliquity.

The nondetection of the secondary eclipse allowed us to place a
constraint on the planet's geometric albedo in the {\it Kepler}
bandpass: $A_g < 0.06$, with 95\% confidence. This is consistent with previous investigations
that showed "hot Jupiters'' often have low albedos
\citep{Esteves2015, Gandolfi2013, Kipping2011}. 

We detected the ellipsoidal light variation and Doppler boosting
effects in the {\it K2} light curve, after filtering out long-term
stellar variability and systematic effects.  The magnitudes of these
two effects imply planetary masses of $2.6 \pm 0.9~M_{\text{Jup}}$ and
$3.9 \pm 1.5~M_{\text{Jup}}$, both of which are consistent with the
mass determined from the spectroscopic Doppler technique \citep{Bryan2012}.  We have
updated the ephemerides of Qatar-2b with the new mid-transit times
observed by {\it K2}.  There is no evidence for orbital decay, leading
to a lower bound on the stellar tidal quality factor $Q'_\star
>1.5\times 10^{4}$~(95\% confidence).

We identified dozens of spot-crossing anomalies in the {\it K2} light
curve. These anomalies revealed the presence of active regions on the
host star along the planet's transit chord. This suggests that
Qatar-2 is magnetically active, as one would expect for a star with
the relatively young age that was determined from gyrochronology.  We
used the observed spot-crossing anomalies to demonstrate that the the
obliquity of Qatar-2 is very likely smaller than 10$^\circ$.  We did
this in two different ways. First we identified individual
spot-crossing anomalies and measured their properties, including their
times of occurrence.  We then used a simple geometric model for which
the parameters were determined by requiring spatial coincidences of
the spot and the planet at the times of observed anomalies.  In a
separate approach, we fitted a photometric model to a portion of the
light curve, based on the premise of a planet transiting a
limb-darkened star with a circular starspot.

Neither model can be relied upon for precise quantitative results,
because of the strong assumptions that were made, such as the circular
shape of the spots and the lack of spot evolution. Nevertheless the
qualitative results leave little room for doubt that the obliquity is
lower than 10$^\circ$. A low obliquity for Qatar-2 is consistent with
a pattern that has been prevoiusly noted: the hot Jupiter hosts with
photospheres cooler than about 6100-6300~K tend to have low
obliquities \citep{Winn2010}.

\bibliography{qatar2}

\begin{table}
\centering
\caption{System Parameters of Qatar-2A}
\begin{tabular}{lrr}
\hline
\hline
{\rm Parameter}  & {\rm }   & {\rm Ref.}  \\
\hline
    Stellar Parameters &$ 	    $&$   $\\
    $T_{\text{eff}} ~({\rm K})$ &$ 4645\pm50     $& A \\
    $\log~g~(\text{dex})$ &$ 	4.601\pm 0.018  $& A \\
    $[\text{Fe/H}]~(\text{dex})$ &$ 	-0.02 \pm 0.08  $& B \\
    $v~\text{sin}~i_\star$ ~(km~s$^{-1})$ &$ 2.8 \pm 0.5   $& A \\
    $M_{\star} ~(M_{\odot})$ &$ 0.74 \pm 0.037     $& A \\
    $R_{\star} ~(R_{\odot})$ &$ 0.713 \pm 0.018     $& A \\
    $\text{Apparent $V$ mag}$ &$ 13.3     $& A \\
    $P_{\text{rot}} ~(\text{days})$ &$ 18.5 \pm 1.9     $& C \\
    Age (Gyrochronology, Gyr) &$ 1.4 \pm 0.3     $& C \\
    Age (Isochrone, Gyr) &$ 15.72 \pm 1.36     $& B \\
     $u_1$ &$ 0.6231 \pm 0.0057     $& C \\
     $u_2$ &$ 0.062 \pm 0.015     $& C \\
    \\
    Planetary Parameters &$ 	    $&$   $\\    

   $P~(\text{days})$ &$ 1.337116553 \pm 0.000000044    $& C \\
   $T_c~(\text{BJD})$ &$ 2455617.581506 \pm 0.000054    $& C \\
    $R_\star/a~$ &$   0.15350 \pm 0.00018 $& C \\
    $a~(\text{AU})$ &$   0.02160 \pm 0.00057  $& C \\
    $R_p/R_*~$ &$   0.16208 \pm 0.00018 $& C \\
    $R_p~(~R_{\text{Jup}})$ &$    1.150 \pm 0.030 $& C \\
    $i~(^{\circ})$ &$   89.7 \pm 0.5 $& C \\
    $b$ &$   0.03 \pm 0.06 $& C \\
    $M_{p, {\rm RV}}~(M_{\text{Jup}})$ &$ 2.487 \pm 0.086     $& A \\
    $M_{p,{\rm ELV}}~(M_{\text{Jup}})$ &$   2.6 \pm 0.9  $& C \\
    $M_{p, {\rm DB}}~(M_{\text{Jup}})$ &$  3.9 \pm 1.5    $& C \\
    $e~({\rm assumed})$  &$  0    $& A \\

\hline
\end{tabular}
\tablecomments{A: \citet{Bryan2012}; B: \citet{Maxted2015}; C: this work.}
\label{tab: para}
\end{table}

\begin{table}
\centering
\caption{Mid-transit Times of Qatar-2b}
\begin{tabular}{lrrr}
\hline
\hline
{\rm Epoch}  & {\rm $T_c$ (BJD~$-$~2454900)}   & {\rm Unc.} & {\rm Ref.}  \\
\hline
$-1197$ &$      717.58156    $& $        0.00082    $ & 1 \\
$-1192$ &$     724.26679    $& $       0.00011   $ & 2\\
$-1141$ &$      792.46109    $& $        0.00270    $ & 1\\
$-1135$ &$      800.47915    $& $        0.00083    $ & 1\\
$-930$ &$    1074.59334    $& $       0.00072    $ & 3\\
$-927$ &$     1078.6048   $& $        0.0012    $ & 4\\
$-921$ &$     1086.62711    $& $        0.00069    $ & 5\\
$-894$ &$    1122.72850    $& $       0.00018    $ & 6\\
$-894$ &$    1122.72800    $& $       0.00017    $ & 7\\
$-894$ &$    1122.72815    $& $       0.00016    $ & 8\\
$-894$ &$    1122.72810    $& $       0.00022    $ & 9\\
$-891$ &$     1126.73842    $& $        0.00033    $ & 10\\
$-888$ &$    1130.7516    $& $       0.0015    $ & 11\\
$-885$ &$     1134.76157    $& $        0.00010    $ & 6\\
$-885$ &$     1134.76196    $& $        0.00012    $ & 7\\
$-885$ &$     1134.76198    $& $        0.00015    $ & 8\\
$-885$ &$     1134.76249    $& $        0.00015    $ & 9\\
$-883$ &$     1137.43785    $& $        0.00068    $ & 12\\
$-882$ &$    1138.77329    $& $       0.00010    $ & 7\\
$-882$ &$    1138.77345    $& $       0.00012   $ & 8\\
$-882$ &$    1138.77312   $& $       0.00015   $ & 9\\
$-877$ &$     1145.45890    $& $        0.00042    $ & 13\\
$-868$ &$    1157.49242    $& $       0.00018    $ & 14\\
$-868$ &$    1157.49246   $& $       0.00028   $ & 15\\
$-868$ &$    1157.49274    $& $       0.00016    $ & 16\\
$-868$ &$    1157.49282    $& $       0.00023   $ & 17\\
$-659$ &$     1436.94926    $& $        0.00059    $ & 18\\
$-657$ &$     1439.62434    $& $        0.00027    $ & 19\\
$-654$ &$    1443.6346   $& $       0.0011    $ & 20\\
$-616$ &$    1494.44559    $& $       0.00056    $ & 21\\
$-604$ &$    1510.4905    $& $       0.0010    $ & 22\\
$-603$ &$     1511.8302    $& $        0.00058    $ & 23\\
$-597$ &$     1519.8502    $& $        0.0013    $ & 24\\
$-580$ &$    1542.58191    $& $       0.00014    $ & 25\\
$0$ &$    2318.109992    $& $       0.000049    $ & 26\\
$1$ &$    2319.447169    $& $       0.000049    $ & 26\\
$2$ &$    2320.784315    $& $       0.000050    $ & 26\\
$3$ &$    2322.121328    $& $       0.000049    $ & 26\\
$4$ &$    2323.458458    $& $       0.000051    $ & 26\\
$5$ &$    2324.795643    $& $       0.000049    $ & 26\\
$6$ &$    2326.132714    $& $       0.000090    $ & 26\\
$7$ &$    2327.469972    $& $       0.000050    $ & 26\\
$8$ &$    2328.807039    $& $       0.000057    $ & 26\\
$9$ &$    2330.143911    $& $       0.000099    $ & 26\\
$10$ &$    2331.481166    $& $       0.000050    $ & 26\\

\hline
\end{tabular}
\tablecomments{ References: (1) Canis Major Observatory
  \citep{Mancini2014}; (2) \citet{Bryan2012}; (3) Strajnic et
  al.~(TRESCA); (4) Zibar M.~(TRESCA); (5) Gonzales J.~(TRESCA); (6)
  MPG/ESO 2.2m $g'$~\citep{Mancini2014}; (7) MPG/ESO 2.2m $r$'
  \citep{Mancini2014}; (8) MPG/ESO 2.2-m $i'$ \citep{Mancini2014}; (9)
  MPG/ESO 2.2m $z'$ \citep{Mancini2014}; (10) Dax T.~(TRESCA); (11)
  Masek M.~(TRESCA); (12) Carreno A.~(TRESCA); (13) Montigiani N.,
  Manucci M.~(TRESCA); (14) Cassini 1.52m \citep{Mancini2014}; (15)
  CAHA 2.2m $g$~\citep{Mancini2014}; (16) CAHA 2.2m $r$
  \citep{Mancini2014}; (17) CAHA 2.2m $z$ \citep{Mancini2014}; (18)
  Campbell J.~(TRESCA); (19) CAHA 1.23m \citep{Mancini2014}; (20)
  Ren'e R.~(TRESCA); (21) Ayiomamitis A.~(TRESCA); (22) Jacobsen
  J.~(TRESCA); (23) Kehusmaa P., Harlingten C.~(TRESCA); (24) Shadic
  S.~(TRESCA); (25) Colazo C., et al.~(TRESCA).; (26) {\it K2} (this
  work). TRESCA stands for ``TRansiting ExoplanetS and CAndidates''.}
\label{tab: ttv}
\end{table}

\begin{table}
\centering
\caption{Table. \ref{tab: ttv} continued}
\begin{tabular}{lrrr}
\hline
\hline

$11$ &$    2332.818338    $& $       0.000052    $ & 26\\
$12$ &$    2334.155345    $& $       0.000059    $ & 26\\
$13$ &$    2335.492644    $& $       0.000079    $ & 26\\
$14$ &$    2336.829644    $& $       0.000096    $ & 26\\
$15$ &$    2338.166710    $& $       0.000060    $ & 26\\
$16$ &$    2339.503878    $& $       0.000056    $ & 26\\
$17$ &$    2340.840863    $& $       0.000056    $ & 26\\
$18$ &$    2342.177965    $& $       0.000071    $ & 26\\
$19$ &$    2343.515114    $& $       0.000049    $ & 26\\
$20$ &$    2344.852294    $& $       0.000055    $ & 26\\
$21$ &$    2346.189436    $& $       0.000068    $ & 26\\
$22$ &$    2347.526675    $& $       0.000090    $ & 26\\
$23$ &$    2348.863781    $& $       0.000075    $ & 26\\
$24$ &$    2350.200849    $& $       0.000049    $ & 26\\
$25$ &$    2351.537860    $& $       0.000052    $ & 26\\
$26$ &$    2352.875083    $& $       0.000068    $ & 26\\
$27$ &$    2354.212212    $& $       0.000052    $ & 26\\
$28$ &$    2355.549382    $& $       0.000053    $ & 26\\
$29$ &$    2356.886369    $& $       0.000052    $ & 26\\
$30$ &$    2358.223494    $& $       0.000054    $ & 26\\
$31$ &$    2359.560744    $& $       0.000049    $ & 26\\
$32$ &$    2360.897787    $& $       0.000059    $ & 26\\
$33$ &$    2362.234890    $& $       0.000054    $ & 26\\
$34$ &$    2363.572009    $& $       0.000055    $ & 26\\
$35$ &$    2364.909290    $& $       0.000106    $ & 26\\
$36$ &$    2366.246114    $& $       0.000051    $ & 26\\
$37$ &$    2367.583411    $& $       0.000049    $ & 26\\
$38$ &$    2368.920242    $& $       0.000078    $ & 26\\
$39$ &$    2370.257464    $& $       0.000054    $ & 26\\
$40$ &$    2371.594669    $& $       0.000056    $ & 26\\
$41$ &$    2372.931863    $& $       0.000048    $ & 26\\
$42$ &$    2374.269011    $& $       0.000055    $ & 26\\
$43$ &$    2375.605990    $& $       0.000056    $ & 26\\
$44$ &$    2376.943205    $& $       0.000051    $ & 26\\
$45$ &$    2378.280327    $& $       0.000069    $ & 26\\
$46$ &$    2379.617357    $& $       0.000051    $ & 26\\
$47$ &$    2380.954439    $& $       0.000054    $ & 26\\
$48$ &$    2382.291665    $& $       0.000071    $ & 26\\
$49$ &$    2383.628818    $& $       0.000079    $ & 26\\
$50$ &$    2384.965898    $& $       0.000053    $ & 26\\
$51$ &$    2386.302902    $& $       0.000052    $ & 26\\
$52$ &$    2387.640011    $& $       0.000091    $ & 26\\
$53$ &$    2388.977140    $& $       0.000051    $ & 26\\
$54$ &$    2390.314349    $& $       0.000050    $ & 26\\
$55$ &$    2391.651335    $& $       0.000051    $ & 26\\
$56$ &$    2392.988456    $& $       0.000058    $ & 26\\
$57$ &$    2394.325632    $& $       0.000048    $ & 26\\
$58$ &$    2395.662874    $& $       0.000052    $ & 26\\

\hline
\end{tabular}
\label{tc2}
\end{table}

\begin{table}
\centering
\caption{Spot-crossing anomalies observed in K2}
\begin{tabular}{lrrrr}
\hline
\hline
{\rm Epoch}  & $t_{\rm anom}$~(BJD~$-$~2454900)   & {\rm Amplitude} & {\rm Width (days)} & {\rm No.} \\
\hline
$      6$ &$          2326.12087 \pm        0.00097    $& $              0.00112 \pm        0.00019   $& $              0.00503 \pm        0.00123    $ & 1 \\
$     18$ &$          2342.15048 \pm        0.00126    $& $              0.00092 \pm        0.00023   $& $              0.00416 \pm        0.00086    $ & 1 \\
$     19$ &$          2343.49928 \pm        0.00059    $& $              0.00154 \pm        0.00021   $& $              0.00336 \pm        0.00049    $ & 1 \\
$     21$ &$          2346.20227 \pm        0.00078    $& $              0.00164 \pm        0.00037   $& $              0.00597 \pm        0.00158    $ & 1 \\
$     22$ &$          2347.55201 \pm        0.00096    $& $              0.00105 \pm        0.00030   $& $              0.00307 \pm        0.00099    $ & 1 \\
$     32$ &$          2360.87726 \pm        0.00041    $& $              0.00195 \pm        0.00025   $& $              0.00296 \pm        0.00048    $ & 1 \\
$     33$ &$          2362.22691 \pm        0.00043    $& $              0.00213 \pm        0.00020   $& $              0.00387 \pm        0.00043    $ & 1 \\
$     34$ &$          2363.57898 \pm        0.00027    $& $              0.00316 \pm        0.00021   $& $              0.00361 \pm        0.00033    $ & 1 \\
$     35$ &$          2364.92923 \pm        0.00118    $& $              0.00192 \pm        0.00034   $& $              0.00808 \pm        0.00187    $ & 1 \\
$     45$ &$          2378.25545 \pm        0.00075    $& $              0.00185 \pm        0.00019   $& $              0.00523 \pm        0.00072    $ & 1 \\
$     46$ &$          2379.60347 \pm        0.00042    $& $              0.00252 \pm        0.00019   $& $              0.00468 \pm        0.00045    $ & 1 \\
$     47$ &$          2380.95384 \pm        0.00035    $& $              0.00299 \pm        0.00019   $& $              0.00484 \pm        0.00038    $ & 1 \\
$     48$ &$          2382.30534 \pm        0.03138    $& $              0.00262 \pm        0.00029   $& $              0.00429 \pm        0.00066    $ & 1 \\
$     49$ &$          2383.65482 \pm        0.00065    $& $              0.00180 \pm        0.00032   $& $              0.00405 \pm        0.00078    $ & 1 \\
$      9$ &$          2330.13165 \pm        0.00169    $& $              0.00152 \pm        0.00028   $& $              0.01160 \pm        0.00234    $ & 2 \\
$     10$ &$          2331.48471 \pm        0.00084    $& $              0.00174 \pm        0.00021   $& $              0.00739 \pm        0.00137    $ & 2 \\
$     11$ &$          2332.83334 \pm        0.02299    $& $              0.00095 \pm        0.00045   $& $              0.00607 \pm        0.00289    $ & 2 \\
$     12$ &$          2334.18152 \pm        0.00097    $& $              0.00092 \pm        0.00046   $& $              0.00243 \pm        0.00152    $ & 2 \\
$     21$ &$          2346.16120 \pm        0.00175    $& $              0.00095 \pm        0.00041   $& $              0.00494 \pm        0.00154    $ & 2 \\
$     22$ &$          2347.50724 \pm        0.00076    $& $              0.00198 \pm        0.00027   $& $              0.00682 \pm        0.00096    $ & 2 \\
$     23$ &$          2348.85870 \pm        0.00098    $& $              0.00318 \pm        0.00089   $& $              0.00958 \pm        0.00205    $ & 2 \\
$     24$ &$          2350.20992 \pm        0.00094    $& $              0.00200 \pm        0.00021   $& $              0.00695 \pm        0.00105    $ & 2 \\
$     25$ &$          2351.55931 \pm        0.00055    $& $              0.00137 \pm        0.00026   $& $              0.00261 \pm        0.00073    $ & 2 \\
$     35$ &$          2364.88900 \pm        0.00115    $& $              0.00187 \pm        0.00035   $& $              0.00577 \pm        0.00195    $ & 2 \\
$     36$ &$          2366.23833 \pm        0.00067    $& $              0.00184 \pm        0.00021   $& $              0.00459 \pm        0.00069    $ & 2 \\
$     37$ &$          2367.58754 \pm        0.00086    $& $              0.00145 \pm        0.00018   $& $              0.00663 \pm        0.00085    $ & 2 \\
$     49$ &$          2383.61220 \pm        0.00081    $& $              0.00156 \pm        0.00031   $& $              0.00514 \pm        0.00129    $ & 2 \\
$     50$ &$          2384.96188 \pm        0.00042    $& $              0.00249 \pm        0.00018   $& $              0.00534 \pm        0.00039    $ & 2 \\
$     51$ &$          2386.31324 \pm        0.00068    $& $              0.00166 \pm        0.00019   $& $              0.00585 \pm        0.00067    $ & 2 \\
$     52$ &$          2387.66170 \pm        0.00065    $& $              0.00144 \pm        0.00022   $& $              0.00356 \pm        0.00075    $ & 2 \\
$     10$ &$          2331.46583 \pm        0.00055    $& $              0.00208 \pm        0.00022   $& $              0.00413 \pm        0.00059    $ & 3 \\
$     11$ &$          2332.81711 \pm        0.00039    $& $              0.00258 \pm        0.00038   $& $              0.00304 \pm        0.00040    $ & 3 \\
$     12$ &$          2334.16835 \pm        0.00039    $& $              0.00244 \pm        0.00023   $& $              0.00380 \pm        0.00040    $ & 3 \\
$     13$ &$          2335.52029 \pm        0.00088    $& $              0.00127 \pm        0.00028   $& $              0.00332 \pm        0.00084    $ & 3 \\
$     23$ &$          2348.84053 \pm        0.00135    $& $              0.00138 \pm        0.00036   $& $              0.00499 \pm        0.00162    $ & 3 \\
$     24$ &$          2350.19119 \pm        0.00100    $& $              0.00160 \pm        0.00022   $& $              0.00504 \pm        0.00123    $ & 3 \\
$     25$ &$          2351.54429 \pm        0.00045    $& $              0.00193 \pm        0.00022   $& $              0.00368 \pm        0.00053    $ & 3 \\
$     26$ &$          2352.89844 \pm        0.00168    $& $              0.00092 \pm        0.00019   $& $              0.00602 \pm        0.00187    $ & 3 \\
$     12$ &$          2334.15237 \pm        0.00053    $& $              0.00138 \pm        0.00026   $& $              0.00273 \pm        0.00055    $ & 4 \\
$     13$ &$          2335.50433 \pm        0.00071    $& $              0.00113 \pm        0.00040   $& $              0.00199 \pm        0.00289    $ & 4 \\
$     21$ &$          2346.18160 \pm        0.00095    $& $              0.00109 \pm        0.00040   $& $              0.00358 \pm        0.00136    $ & 5 \\
$     22$ &$          2347.53394 \pm        0.00141    $& $              0.00076 \pm        0.00041   $& $              0.00281 \pm        0.00397    $ & 5 \\
$     23$ &$          2348.88640 \pm        0.00076    $& $              0.00166 \pm        0.00051   $& $              0.00346 \pm        0.00100    $ & 5 \\
$     35$ &$          2364.90513 \pm        0.00032    $& $              0.00217 \pm        0.00033   $& $              0.00189 \pm        0.00060    $ & 5 \\
$     49$ &$          2383.63367 \pm        0.00098    $& $              0.00099 \pm        0.00036   $& $              0.00395 \pm        0.00168    $ & 5 \\
$     52$ &$          2387.62708 \pm        0.00040    $& $              0.00180 \pm        0.00026   $& $              0.00254 \pm        0.00039    $ & 6 \\
$     53$ &$          2388.97723 \pm        0.00098    $& $              0.00147 \pm        0.00023   $& $              0.00952 \pm        0.00237    $ & 6 \\
$     54$ &$          2390.33064 \pm        0.00092    $& $              0.00105 \pm        0.00023   $& $              0.00350 \pm        0.00091    $ & 6 \\

\hline
\end{tabular}

\label{tab: anomalies}
\end{table}

\begin{table}
\centering
\caption{Results of the geometric model}
\begin{tabular}{lrrrr}
\hline
\hline
{\rm Parameter}  & {\rm Spot 1}   & {\rm Spot 2} & {\rm Spot 3} & {\rm Combined} \\
\hline
$     \lambda~(^{\circ})$&$          0.0 \pm        4.4    $& $              0.0 \pm        4.4   $& $              0.0 \pm        6.3    $ & $              0.0 \pm        2.8    $ \\
$     i_s~(^{\circ}) $ &$          90 \pm        22    $& $              90 \pm        22   $& $              90 \pm        24    $ & $              90 \pm        19    $ \\
$     l~(^{\circ})$ &$          1 \pm        18    $& $              2 \pm         19   $& $              1 \pm        21    $ & $              2 \pm        16    $ \\
$     P_{\text{rot}}~(\text{days})$ &$          17.89 \pm        0.14    $& $              17.98 \pm        0.15   $& $              17.99 \pm        0.50    $ & $              17.94 \pm        0.10    $ \\
\hline
\end{tabular}

\label{tab: geometric}
\end{table}

\begin{table}
\centering
\caption{Results of the Numerical models}
\begin{tabular}{lrr}
\hline
\hline
{\rm Parameter}  & {\rm {\tt Spotrod}}   & {\rm Pixelated Model} \\
\hline
$     \lambda~(^{\circ})$&$          1.4^{+3.0}_{-1.7}    $& $              4.3^{+4.6}_{-2.7}  $ \\
$     i_s~(^{\circ}) $ &$          89.8 \pm        4.0    $& $              90.1 \pm        7.9   $ \\
$     l~(^{\circ})$ &$          6.2 ^{+7.3}_{-4.2}           $& $              5.4 ^{+7.5}_{-6.8}   $\\
$     P_{\text{rot}}~(\text{days})$ &$          18.62 \pm        0.31    $& $              18.54 \pm        0.40   $ \\
$     t_{\rm anom}~(\text{BJD~$-$~2454900})$ &$          2380.953 \pm  0.025         $& $              2380.928 \pm        0.025   $ \\
$     \alpha ~(^{\circ}) $ &$         11.2  \pm   1.2         $& $              12.7 \pm        0.7   $ \\
$     f$ &$          0.90 \pm        0.02    $& $              0.92 \pm        0.01   $ \\
\hline
\end{tabular}

\label{tab: numerical}
\end{table}

\end{document}